\documentclass{article}
\begin{document}
\title{A Doppler-like strong light-matter interaction.
\author{J. Moret-Bailly 
\footnote{Laboratoire de physique, Université de Bourgogne, BP 47870, F-21078 
Dijon cedex, France.
email : jmb@jupiter.u-bourgogne.fr
}}}
\maketitle

\begin{abstract}
While complicated, unreliable alternatives to Doppler effect were proposed, an elementary optical light-
matter interaction provides one which is commonly observed in the labs, but with a distortion due to the 
use of short, powerful laser pulses.

It is generally assumed that Raman scattering in gases is incoherent. This assumption fails if the 
pressure is lowered enough to increase the relaxation times over the length of light pulses;  the 
''Impulsive Stimulated Raman Scattering'' (ISRS), generally used to study dense matter with 
ultrashort laser pulses, is adapted to the low energy pulses making the incoherent light beams; the 
usual light is redshifted by some very low pressure gases while it propagates.

To produce this adapted ISRS called ''Incoherent Light Spatially Coherent Raman Scattering" 
(ILSCRS), a molecule must have an hyperfine structure: polyatomic molecules must be heavy or 
have odd numbers of electrons; light atoms and the other molecules must be perturbed by a Stark 
or Zeeman effect.

ILSCRS redshifts may be distinguished from Doppler redshifts using a very difficult to observe 
dispersion of ILSCRS redshifts. This dispersion may explain the discrepancies of the fine structures 
in the spectra of the quasars, presently attributed to a variation of the fine structure constant.

While the present interpretation of the Lyman forest in the spectra of quasars requires clouds 
stressed, for instance, by sheets of dark matter, ILSCRS interpretation requires only usual physical 
concepts. It produces thermal radiations from short wavelengths, just as dust.


\end{abstract}

\maketitle
%

\section{Introduction}
Twenty years ago, many astrophysicists thought that Doppler effect was unable to explain some observed 
redshifts. But they were unable to find a credible alternative because they introduced two restrictions:

- They generally supposed that the electromagnetic field is sinusoidal. A consequence of this hypothesis is 
that if the received frequency is not equal to the emitted frequency, the number of wavelengths between the 
source and the receiver changes permanently, that is the distance between them changes, the frequency 
shift is a Doppler shift. Thus, looking for a non-Doppler frequency shift which would help the interpretation 
of some astrophysical data, Marmet \cite{Marmet} wrote that it is necessary to consider pulsed light; but 
the process he proposed is not convincing.

- They considered that individual molecules were involved in the interactions, while the molecules must be 
packed into sets in the theory of many optical effects, refraction, amplification in a laser,...

Unaware of astrophysics, working in pure optics, it appeared to us that an interaction of ordinary 
incoherent light with gases induces redshifts which may be confused with Doppler redshifts 
\cite{Moret,Moret1,Moret2}; this interaction, although strong, seemed unobservable in the labs, requiring 
extremely low pressure gases, thus very long optical paths. This effect appears now as an avatar of the 
"Impulsive Stimulated Raman Scattering" (ISRS), commonly observed in the labs using ultrashort laser 
pulses.

\medskip

This paper has two aims: - first, compared to the previous ones, it is a much more precise and, I hope, 
easier to read description of the new effect called "Incoherent Light, Spatially Coherent Raman Scattering" 
(ILSCRS); a short summary of the sections, at their beginnings, allows the reader to have an overview 
sufficient to follow the paper without studying the following demonstrations. - second, it gives the 
correspondence between ISRS and ILSCRS, setting precisely what happens when the short, powerful 
laser pulses replace the longer, weaker pulses which make the natural incoherent light. We hope that this 
connection between ISRS and ILSCRS will convince the astrophysicists of the reality of ILSCRS, so that 
they take it into account in their models.

\medskip

	The properties of pulsed light have been extensively studied using ultrashort, femtosecond laser 
pulses; compared to laser pulses, the pulses of incoherent light are longer by a factor of the order of 
$10^5$; the definition of an ''ultrashort pulse'' was given by G. L. Lamb as ''shorter than all relevant time 
constants''\cite{Lamb}; in a gas, for a Raman effect, these time constants are the time between collisions 
which may be increased by a decrease of pressure , and the period corresponding to the Raman transition 
which may be chosen long enough. Thus, the nanosecond pulses whose ordinary incoherent light is made 
of, are \textquotedblleft ultrashort''.

ISRS, described by Yan et al. \cite{Yan}, is mostly used to study fast evolutions of dense matter 
\cite{Eichler,Salcedo, Nelson1,Nelson2,Robinson,De Silvestri}.
 The frequency shift is the result of an interference of the exciting beam with a Raman scattered beam. The 
emission of the scattered light is stimulated by the powerful exciting beam, so that the scattered amplitude 
is proportional to the square of the exciting field; thus the frequency shift depends on the intensity of the 
light; in ILSCRS the Raman scattered field is spontaneous, proportional to the exciting field, so that the 
frequency shift does not depend on the exciting field.

\medskip
The summaries of the subsections are followed by more technical explanations and demonstrations, made 
as simple as possible using the semiclassical theory, following Bloembergen \cite{Bloembergen}: 
\textquotedblleft The subtle interplay between real and imaginary parts of the complex linear and nonlinear 
susceptibilities follows quite naturally from the semiclassical treatment. [...] The semiclassical theory which 
is used in this monograph will describe all situations correctly in a much simpler fashion\textquotedblright. 

\medskip
Section 2 recalls the general properties of ISRS and adapts them to ILSCRS.

Section 3 gives some spectroscopic properties of gases subject to ILSCRS.

Section 4 suggests possible applications of ILSCRS in astrophysics: we do not intend to give reliable 
interpretations of astrophysical observations.

\section{ Properties of ISRS and ILSCRS}
This section presents the properties of ISRS and the conditions which must be fulfilled to preserve these 
properties replacing short powerful laser pulses by incoherent light, getting ILSCRS.
%

\subsection{Space coherence}
{\it With space coherence, the output wave surfaces belong to the family of the input wave surfaces, so 
that the images are sharp, without any blur. To get the space coherence, nearly no collision must happen 
during a light pulse; in ILSCRS, only low pressure gases work.}

\medskip
In the Huygens' construction of the wave surfaces, each point of a particular wave surface is considered as 
a source; the envelop of the wavelets produced by these sources after a short time $\Delta t$ is new wave 
surface. This construction may be extended replacing the Huygens' sources by scattering molecules, but 
some modifications and conditions appear:

- the number of molecules, thus the number of scattered waves is not infinite, so that the building of a new 
wave surface is not perfect. The most important consequence is the incoherent Rayleigh scattering (blue of 
the sky) which perturbs the refraction; we will consider that the fluctuations of the molecular density are 
low, so that such effects may be neglected.

- the phases of the waves scattered by all molecules lying on a particular wave surface must be the same; 
the classical and quantum computations of the scattered waves are identical for all molecules, unless 
collisions introduce phase changes.

\medskip
To obtain the space coherence, during a light pulse it must be almost no collision between the molecules. 
The mean time between two collisions in a gas of identical spherical hard molecules is:
\begin{equation}
\tau=\frac{1}{2Nd^2}\sqrt{ \frac{m}{4\pi kT}} \label{temps}
\end{equation}
 where $N$ is the number of molecules by unit of volume, $d$ their diameter, $m$ their mass and $T$ the 
temperature. This mean time gives only an order of magnitude because the molecules are not hard 
particles, so that it is difficult to define a length considered as the diameter of a molecule.

\medskip

In conventional Raman spectroscopy, using regular sources, for instance mercury vapour lamps, the 
Raman intensity is so low that the pressure of a studied scattering gas must be of the order of magnitude of 
the atmospheric pressure; as the pressure in the source (made of heavy atoms) is lower than the pressure 
in the studied gas, the mean time between collisions is shorter than the duration of the pulses of light (about 
10 nanoseconds), so that conventional Raman scattering is incoherent. 

The incoherence is widely responsible of the weakness of the conventional Raman scattering: : Set $a \exp 
(i \phi_j)$ the complex amplitude scattered into a point by a molecule number $j$ ($j=1\dots n $, where n 
is the number of scattering molecules). The total scattered intensity is:
\begin{equation}
I=\Bigl( \sum_ j a \exp (i \phi_j) \Bigr) \Bigl( \sum_k a^* \exp(-i\phi_k)\Bigr).
\end{equation}
In an incoherent scattering process, the $\phi_j$ are stochastic so that the mean value of 
$\exp(i\phi_j)\exp(i\phi_k)$ is zero if $j$ is not equal to $k$, else one; thus the total intensity is 
$I_I=naa^*$. In a coherent scattering process, supposing that the wave surface converges into an usual 
diffraction figure, at the centre of this figure all $\phi_j$ are equal, so that $I$ takes the large value 
$I_C=n^2aa^*$; out of the diffraction figure, the scattered amplitudes cancel if the fluctuations in the 
repartition of the molecules are neglected. In conclusion, the incoherent process scatters a very low 
intensity in all directions; the coherent process scatters a strong intensity which produces, if the wavelength 
are nearly the same, the same diffraction pattern than the incident beam. As in astrophysics, we will always 
consider wide beams, so that diffraction may be neglected.

In many places of the universe, the pressure is so low that the pulses of ordinary light, long of 10 
nanosecond or less, may be said \textquotedblleft ultrashort\textquotedblright by a comparison with the 
mean time between collisions; in these places, Raman scattering is space coherent.

\subsection{Interference of the incident and scattered waves into a single wave}
{\it Consider a single Raman transition. In conventional or coherent Raman scattering, the frequency of the 
scattered light gets a shift corresponding to a molecular transition; for ISRS or ILSCRS the period 
corresponding to the Raman transition must be larger than the length of the pulses, so that the two space-
coherent beams interfere into a single monochromatic beam whose frequency is intermediate; the 
frequency shift is proportional to the regular Raman shift and to the scattered amplitude; the width of the 
exciting line is not increased.

With incoherent light, the Raman transition must correspond to a radio-frequency, the gas must have 
hyperfine (or equivalent) populated levels.}

\medskip
In the theory of refraction, the incident field induces in each molecule a dipole which radiates a wave 
dephased of $\pi/2$. The incident and scattered light interfere into a single wave, generally late in 
comparison with the incident wave.

In the semi-classical theory of the Raman effect, the dipole induced by the incident field is coupled with the 
dipole which radiates the Raman wave; at the beginning of a light pulse, the dipoles are dephased of 
$\pi/2$, so that the phases of the incident and scattered fields are the same, modulo $\pi$. During the 
pulse, the phase changes because the dipoles have different frequencies.

The interferences of two different frequencies is often observed, for instance between the two beams of a 
Michelson interferometer, when one of the mirrors moves, producing a Doppler frequency shift. Show by 
an elementary computation that, if this phaseshift is lower enough than $\pi$, the sum of the incident and 
scattered fields is a single field having an intermediate frequency

The electric field in a pulse of light is the product of a slow varying electric field $E(t)$ giving the pulse 
shape by a sine function; for an exciting field of frequency $\nu_e$ the field may be written $E(t) \cos (2\pi 
\nu_et)$ and a field scattered at a frequency $\nu_s$ by a thin layer of thickness $L$ of gas, with the same 
polarisation and the same phase at the beginning of the pulse ($t=0$) :
$E(t)Lq \cos(2\pi  \nu _st)$ , where the product $Lq$ is a small dimensionless coefficient including the 
thickness $L$; $Lq$ will be a first order quantity; the sum of the two emerging fields is:
\begin{equation}
\begin{array}{l}
D = E(t) (1-Lq)\cos(2\pi  \nu _e t)+E(t)Lq\cos(2\pi  \nu _st)=\\
=E(t)(1-Lq)\cos (2\pi  \nu _e t)+\\
+E(t)Lq(\cos (2\pi  \nu _et)\cos (2\pi (\nu _s - \nu _e)t)-
\sin (2\pi  \nu _et)\sin (2\pi (\nu _s - \nu _e)t))\,.
\end{array}
\end{equation}
Writing the Raman frequency $ \nu_i =\nu_s - \nu_e$ and the length of the pulse $t_0$, suppose
\begin{equation}
| \nu_i t_0 |<<\pi,\,\label{condi}
\end{equation}
that is the Raman period is much larger than the length of the pulses; we may develop the corresponding 
trigonometric functions:
\begin{equation}
\begin{array}{l}
D \approx E(t) \cos (2\pi  \nu _et)-2E(t)Lq\pi (\nu _s - \nu _e)t\sin (2\pi  \nu _et)\\
+2E(t)Lq(\pi (\nu _s - \nu _e)t)^2\cos(2\pi  \nu _et)
+(4/3)E(t)Lq(\pi (\nu _s - \nu _e)t)^3\sin (2\pi  \nu _et)\,.
\end{array}
\end{equation}
set:
\begin{equation}
\tan (\phi t)=2Lq\pi (\nu_s - \nu_e)t=2Lq\pi \nu_it;\:(-\pi /2<\phi \leq \pi/2)\,;\label{eqq}
\end{equation}
\begin{equation}
\phi \approx 2Lq\pi \nu_i \label{phi}
\end{equation}
is a first order quantity.
\begin{equation}
\begin{array}{l}
D \approx E(t) (\cos (2\pi  \nu _et)\cos( \phi t)-\sin (2\pi  \nu _et)\sin (\phi t))
/\cos(\phi t)+\\
+2E(t)Lq(\pi (\nu _s - \nu _e)t)^2\cos(2\pi  \nu _et)
+(4/3)E(t)Lq(\pi (\nu _s - \nu _e)t)^3 \sin (2\pi  \nu _et)\,.
\end{array}
\end{equation}

In a first order approximation :
\begin{equation}
D \approx E(t) \cos ((2\pi \nu _e +\phi )t).\label{eqs}
\end{equation}
The waves interfere into a single wave within the pulse.  Thus, in place of the emergence of a new line 
shifted of $\nu_i $ , the whole incident flux gets the slight frequency shift 
\begin{equation}
\Delta\nu = \phi/2\pi = Lq\nu_i.\label{delnu}
\end{equation}

\medskip
Using wide beams to neglect the diffraction, the intensity of the conventional coherent Raman effect is 
limited because the difference of the indices of refraction for the incident and scattered frequencies 
introduces a phase-shift of the beams scattered on two different wave surfaces: the beams scattered by far 
wave surfaces have various phases, their interference is partly destructive. As in ISRS there is a single 
frequency, a single beam, there is no limitation of the effect by the dispersion of the refraction.

\medskip
With ordinary incoherent light, the period corresponding to the Raman transition must be longer than 10 
nanoseconds: the molecule must have Raman transitions in the radio-frequencies domain, in practice the 
molecule must have an hyperfine structure.

Heavy molecules and atoms have a convenient hyperfine structure, but they are probably not abundant in 
the gaseous part of the universe. The nuclear spin hyperfine structures, for instance in $H_2$, are not 
efficient; on the contrary, light polyatomic molecules having an odd number of electrons have hyperfine 
structures whose interactions with light are strong.

Hyperfine structures are induced in all molecules by Stark and Zeeman effects; the width of the structure, is 
often proportional to the field; if it is not too large, so that condition (\ref{condi}) remains true, the 
frequency shift is proportional to the square of the field.
\subsection{Linearity of ILSCRS}
{\it In ISRS, the scattered amplitude is proportional to the square of the incident amplitude because it is 
stimulated by the exciting laser beam, so that the redshift is proportional to the intensity of the light beam 
and a complex spectrum is distorted. In ILSCRS, the spontaneously scattered amplitude is proportional to 
the incident amplitude, so that the frequency shift does not depend on the intensity, the spectra are shifted 
without distorsion.}

It seems to be a discrepancy for the amplitudes scattered in ISRS and ILSCRS. But in quantum 
electrodynamics, and in the correct semi-classical theory, a spontaneous emission is an amplification of the 
corresponding mode of the zero point field. Thus the electric field of the exciting beam is the sum of a zero 
point field $\bf E_0$ and the field $\bf E$ of the old classical theory. The computation of ISRS shows that 
the scattered amplitude is proportional to $\bf (E_0+E)^2$ that is to  $(E_0+E)^2$ because the 
amplification does not change the polarisation. As the power is large in ISRS, $E_0$ may be neglected, 
the scattered amplitude is proportional to the square of the field defined either in the old or in the recent 
theories. On the contrary, in natural light $E$ is much smaller than $E_0$ so that $E^2$ may be neglected; 
$E_0$, the intensity of the zero point field is always cancelled; it remains $2E_0E$ , proportional to $E$ 
because, in the average, $E_0$ is constant.

There is no threshold, a well known property of ISRS. To obtain a redshift independent of the intensity in 
the laboratory, using short laser pulses, the beams must be enlarged and the peak intensity decreased to 
get an electric field only a little larger than the zero point field.
\subsection{Elementary computation of the ILSCRS redshift.}
{\it A complete computation of the ILSCRS redshift requires the knowledge of the traces of the tensors of 
polarizability for all radio-frequency Raman transitions, for all exciting frequencies. For the molecules 
having an odd number of electrons, chemically reactive, these data are difficult to measure or compute. In 
a classical representation of the Raman excitation of a molecule, we assume that the excited dipole is 
strongly coupled to the radiating dipole, so that the ILSCRS scattering has the same order of magnitude 
than the scattering which produces the refraction.}

An exciting electromagnetic field induces for each Raman transition $i$ a scattered field proportional to an 
element of the tensor of polarizability; averaging this result for all orientations of the gaseous molecules, the 
scattered field is proportional to the trace $\beta_i$ of the tensor of polarizability and has the same 
polarisation than the exciting field; thus, the electric field may be considered as a scalar.

The amplitude ${p_i}$ of the dipole induced in an unit volume for a transition $i$ is proportional to the 
incident electrical field and to the number $N_i$ of molecules per unit of volume in the convenient state: 
$p_i=N_i\beta_i E$. The field scattered at the exciting frequency $\nu_e$ produces the refraction through 
$ P, N$ and $\alpha$ similar to $p_i, N_i$ and $\beta _i$ .

At the thermal equilibrium, $N_i$ is deduced from a Boltzman factor $B_i$, so that the ratio of a Raman 
dipole, with respect to the refracting dipole is:
\begin{equation}
\frac{p_i}{P}=\frac{N_i\beta_i}{N\alpha}=\frac{B_i\beta_i}{\alpha};
\end{equation}
The ratios of scattered amplitudes are the same for a single molecule to any direction, or for a large set of 
identical molecules on an exciting wave surface to the initial direction of propagation\footnote{The relation 
between the coefficients in the two configurations requires the addition of Huygens\textquoteright\hskip 
1mm wavelets by a simple but tedious integration called \textquotedblleft the optical 
theorem\textquotedblright.}.

\begin{equation}
\frac{q_i}{Q}=\frac{p_i}{P}=\frac{B_i\beta _i}{\alpha}. \label{rap}
\end{equation}

\medskip
Recall the elementary theory of the refraction index $n$ with our notations:
Set $E\cos(2\pi \nu t)$ the electric field of a wave of frequency $\nu$ as it reaches a thin sheet of gas 
whose thickness $L$ is a first order small quantity; the absorption through this sheet is neglected so that 
the output field is $E\cos(2\pi \nu (t-L/c))$; the Rayleigh scattered field being late of $\pi /2$, the total 
output field is:
\begin{equation}
E[\cos 2\pi\nu(t-L/c)+LQ \sin 2\pi\nu (t-L/c)]
\approx E \cos [2\pi\nu(t-L/c)-LQ]
\end{equation}
The refraction index $n$ is obtained by an identification of this field with $E \cos 2\pi\nu (t-nL/c)$, giving 
$n-1=cQ/(2\pi\nu)$.

The dynamical dielectric constant $\epsilon$ which equals $1+4\pi N\alpha$ is nearly 1 in a dilute gas, so 
that its square root $n$ equals $1+2\pi N\alpha$; thus:
\begin{equation}
n-1 = 2\pi N\alpha =cQ/(2\pi \nu).\label{ind}
\end{equation}
By equations \ref{rap} and \ref{ind}:
\begin{equation}
Lq_i=LQB_i\beta _i/\alpha =4\pi^2NB_i\beta_i\nu L/c\,.
\end{equation}
From equation \ref{delnu}, we get the shift:
\begin{equation}
\Delta \nu=\sum_i[\nu_iLq_i]=4\pi^2N\sum_i[B_i\beta_i\nu_i]\nu L/c\,.\label{eqa}
\end{equation}
In a first order development $B_i$ is proportional to $\nu_i$, so that the contribution of transition $i$ 
to the lineshift is proportional to $\nu_i^2$.

Consider \cite{Moret,Moret1} a model molecule having a high energy level and two low lying levels of 
energies $E_1$ and $E_2$, close enough to allow a series development of the exponent in the Boltzman 
factor, so that the difference of the populations in the low states is:
\begin{equation}
N_1-N_2 = N(B_1-B_2) = N\Bigl[\exp \Bigl (\frac{-E_1}{2kT} \Bigr)- \exp \Bigl(\frac{-
E_2}{2kT}\Bigr) \Bigr] \approx N\frac{E_2-E_1}{2kT}= N\frac{h\nu_i}{2kT}.\label{Bol}
\end{equation}
 $\beta_i$ is different from $\alpha$ because it requires a transfer of energy from the excited oscillator 
to the radiating one; but this transfer is generally fast, so that generally $\beta_i$ and $\alpha$ have the 
same order of magnitude; thus, assume the rough approximation $\alpha = 2\beta_1 = 2\beta_2$. 
Equation \ref{rap} becomes $q_i = B_iQ/2 = -QE_i/(4kT)$ for $i =$ 1 or 2. From equations 
\ref{delnu}, then \ref{ind} the frequency shift is
\begin{equation}
\Delta\nu = L (q_1-q_2) \frac {E_1-E_2}{h} = -\frac{L}{4hkT}( E_1-E_2)^2  \frac{\pi\nu}{c}(n-1)
= \frac{Lh\nu_i^2}{2kT} \frac{\pi\nu}{c}(n-1). \label{dafin}
\end{equation}

Replace the index of refraction using the classical dispersion formula for a single spinless electron 
resonating at the high frequency $F$ which correspond to a transition between a low and the upper 
molecular states :
\begin{equation}
\frac{\Delta\nu}{\nu} = \frac{\pi L( E_1-E_2)^2}{2hckT}\frac{Ne^2}{8 \pi^2m\epsilon_0F^2}=
\frac{ Lh}{2ckT}\frac{Ne^2}{8 \pi m\epsilon_0}\Bigl(\frac{\nu_i}{F}\Bigr)^2 \label{dfin}
\end{equation}
where $\nu_i=\nu_1= ( E_1-E_2)/h$ is the regular Raman frequency shift, $e$ and $m$ are the charge 
and the mass of the electron.

\subsection{To an experimental measure of the parameters of ILSCRS}
As the hyperfine structures are generally complex, it seems difficult to compute the parameters of ILSCRS, 
except, maybe for simple but important molecules such as $H_2^+$. Is an experimental measure possible 
?

The Raman transitions which produce ILSCRS or ISRS are selected by the length of the pulses; thus, to 
get the parameters of ILSCRS redshift by a single experiment, it is necessary to use either incoherent light, 
or laser pulses whose lengths reproduce the lengths of the pulses in natural light.

Some active molecules, such as NO are stable, but it seems difficult to build a low pressure multiple path 
cell long enough to obtain a measurable ILSCRS redshift. It appears much easier to measure ISRS 
redshifts, even using long pulses. The correlation between ILSCRS and ISRS which was shown in 
subsection 2.3 gives the parameters of ILSCRS from those of ISRS, using long pulses.

\subsection{Dispersion}

{\it In a first approximation, the dispersion of the tensor of polarisability is neglected, so that the relative 
frequency shift $\Delta\nu/\nu$ does not depend on the frequency. Introducing the dispersion, it may be 
possible, but uneasy, to distinguish an ILSCRS redshift from a Doppler redshift.}

The $\alpha$ and $\beta_i$ in equation \ref{eqa} are computed from tensors of polarizability which, in a 
first approximation, do not depend on the frequency of the exciting light. 
If the frequency of an exciting beam is nearly the frequency of a dipolar absorption, Raman scattering is 
said  \textquotedblleft resonant'', the scattered amplitude is large. A polyatomic molecule has a large 
number of relatively weak absorption lines, thus a large number of relatively weak ILSCRS resonances: it 
will be difficult to observe them. On the contrary, the atomic lines are strong and well separated in the 
spectrum: the detection of ILSCRS resonances will be easier on light atoms perturbed by a Zeeman or 
Stark effect.

\medskip
Using natural light, a low pressure gas, the de-excitation of the gas pumped by a redshifting ILSCRS 
process cannot result from collisions; but, in radio-frequencies, the spontaneous emissions of light are 
weak. Thus, the ILSCRS redshifting process is limited by the molecular de-excitation processes which 
may be an amplification of low frequency electromagnetic field (which includes the spontaneous emission, 
considering the zero-point field), an ILSCRS blueshift of a low temperature\footnote{The temperature of 
an optical mode is defined from the Planck's laws.} optical mode \dots

These simultaneous processes must be bound into a parametric interaction\footnote{ILSCRS and ISRS 
describe only a part of the parametric process; a good acronym would be too long !}: the active molecule 
appears as a sort of catalyst which helps a flood of energy from the hot modes which are redshifted to the 
cold modes. As the intensity of the parametric interaction is limited by its weakest component, that is the 
transfer of energy to the cold modes which does not depend on the frequency of the hot modes, the 
dispersion of the redshift is reduced.
\section{Applications of ILSCRS in low pressure gases}
This section presents consequences of ILSCRS interactions in low pressure gases.

\subsection{Visibility of lines absorbed by an ILSCRS active gas}
{\it The linewidth of an absorption or emission line of a red-shifting gas equals the redshift: thus the line is 
weak, wide, it cannot be observed individually; the absorption by a set of lines may be confused with an 
absorption by dust.}

\medskip
The absorption of the gas, supposed homogenous, is generally represented by an absorption coefficient 
$k(\nu)$ verifying:
\begin{equation}
{\rm d} \Phi (\nu, x)/ \Phi (\nu, x ) = k(\nu){\rm d} x
\end{equation}

where $\Phi (\nu, x)$ is the flux of energy of a light beam of frequency $\nu$ propagating along an $Ox$ 
axis.

	Consider a light flux initially in a small spectral range near a frequency $\nu_0$. If there is, along 
the light path, a redshift having any origin, expansion of the universe or ILSCRS, the frequency of this flux 
scans the whole redshift; its local absorption is a function of its local frequency $\nu_x$; the total 
absorption from $x=0$ to $x=X$ is

\begin{equation}
\Delta \Phi =\int_0^X{k(\nu_x){\rm d} x}
\end{equation}

As $\nu_x$ varies continuously in the range of the redshift, the function $ k(\nu_x)$ scans the spectrum; 
the sharpest lines get a width larger than the redshift : they cannot be observed individually; as they 
generally obscure the high frequencies more than the low frequencies, in a partial observation of a 
spectrum, their absorption may be confused with an absorption by dust. In a region where the density of 
electromagnetic energy is large, the correlated increase of thermal radiation by ILSCRS may seem due to 
an increase of the temperature of the dust.
\subsection{Detection of a red-shifting gas }
{\it Resonances in a red-shifting gas introduce slight variations of the relative frequency shift of the lines of a 
spectrum; if the spectrum is rich, the positions of the resonances may be found with a precision sufficient to 
characterise the red-shifting gas.

Discrepancies observed in spectra of quasars may be ILSCRS signatures, rather than a consequence of a 
variation of the fine structure constant.}

\medskip
The frequency shift may be written:
\begin{equation}
{\rm d}  \nu = \frac{\rho\nu}{q}{\rm d} x
\end{equation}
where $\rho$ is the density of the gas and $q$ a parameter depending on the gas. Supposing that the 
composition of the gas is constant, the equation is integrated from the frequencies $\nu_i$ of lines emitted 
or absorbed at $x=0$ to the frequencies $\nu_{io}$ observed at $x=X$ :
\begin{equation}
\int_0^X\rho{\rm d} x=\int_{\nu_i}^{\nu_{io}}q\frac{{\rm d} \nu}{\nu}       
=\int_{\nu_j}^{\nu_{jo}}q\frac{{\rm d}  \nu}{\nu}=\dots\label{dispersion}
\end{equation}
To explicit that q depends slightly on $\nu$, set $q=q_0+r(\nu)$ where $q_0$ is a constant, and the 
average of the small function $r(\nu)$ is zero. From equation \ref{dispersion}:
\begin{equation}
q_0\bigl[\ln\bigl(\frac{\nu_i}{\nu_{io}}\bigr)- 
\ln\bigl(\frac{\nu_j}{\nu_{jo}}\bigr)\bigr]=\int_{\nu_i}^{\nu_{io}} r(\nu)\frac{{\rm d} \nu}{\nu}-
\int_{\nu_j}^{\nu_{jo}} r(\nu)\frac{{\rm d} \nu}{\nu}= \int_{\nu_i}^{\nu_j} r(\nu)\frac{{\rm d} 
\nu}{\nu}-\int_{\nu_{io}}^{\nu_{jo}} r(\nu)\frac{{\rm d} \nu}{\nu}\label{disp2}
\end{equation}
Supposing that $\nu_i-\nu_j$ is small, $r(\nu)$ may be replaced by its mean value $m(\nu_i,\nu_j)$ 
between $\nu_i$ and $\nu_j$, so that equation \ref{disp2} becomes:
\begin{equation}
 \ln\bigl(\frac{\nu_i}{\nu_{io}}\bigr)- \ln\bigl(\frac{\nu_j}{\nu_{jo}}\bigr) =\frac{2(\nu_j-\nu_i) 
m(\nu_i,\nu_j)}{q_0(\nu_i+\nu_j)} -\frac{2(\nu_{jo}-\nu_{io}) 
m(\nu_{io},\nu_{jo})}{q_0(\nu_{io}+\nu_{jo})}\label{disp3}
\end{equation}
Supposing that the frequency shift is purely Doppler, the first member of equation \ref{disp3}  is zero; 
Webb et al. \cite{Webb} observed that it is non-zero and wrote that it is due to a variation of the fine 
structure constant. ILSCRS seems a simpler explanation. As all frequencies are known, equation 
\ref{disp3} gives a relation between $ m(\nu_i,\nu_j)/q_0$ and $ m(\nu_{io},\nu_{jo})/q_0$.

To study the resonances which appear if absorption lines lie on the intervals  $(\nu_i,\nu_j)$ {\it or} $ 
(\nu_{io},\nu_{jo})$, it is necessary to distinguish what comes from one or the other interval. This seems 
possible by a statistical study if lots of lines are measured. The quasars spectroscopy seem particularly 
favourable because:

- observing similar quasars, sharp absorption lines are observed with many different redshift, so that the 
averages $ m(\nu_i,\nu_j)$ and $ m(\nu_{io},\nu_{jo})$ are local and significant.

-  the active gases are made of atoms, so that there are few absorption lines, thus few, separated 
resonances;

- these resonances  are near known dipolar absorption lines, so that their observation will be a test of 
the appropriateness of the theory for the quasars. Unhappily, up to now, we could not get the good 
spectra which show discrepancies in the hypothesis of pure Doppler effect.


\subsection{Multiplication of absorption lines in a variable magnetic field}
{\it Light atoms do not have strong hyperfine transitions: they do not redshift the light, so that their 
absorption lines are visible; but if there is an electric or magnetic field, the redshift appears, the absorption 
disappears. Thus, for a light ray propagating in a variable magnetic field, lines are written as pseudo-lines 
into the light where the field is nearly zero, shifted when the field appears, written into an other place for 
another zero of the field, \dots Each pseudo-line is sharp, although it has strong feet. }

\medskip
Consider a plasma of atoms at a temperature of 10 000K, with low enough a pressure to produce 
ILSCRS redshifts.

\medskip
The ratio of the populations in two levels defining a low energy Raman transition, $\exp(h\nu_i /kT)$ at the 
equilibrium, may be developed, because the temperature is high; thus, the difference of $B_i$ for a Stokes 
and its antiStokes interaction is proportional to $h\nu_i/kT$, so that, by eq.\ref{eqa}, $\Delta \nu/\nu$ is 
proportional to $\sum_i[\beta_i\nu_i^2]$.

Usually, the $\nu_i$ are proportional to a field $H$ ; define a variable $x$ proportional to the mass of gas 
passed by the light beam through a unit of surface. As the frequency shift is proportional to the squares of 
the $\nu_i$, thus to the square of the magnetic field, we may write:
\begin{equation}
{\rm d} \nu =A H^2(x){\rm d}x\, ,\label{eqb}
\end{equation}
where $A$ depends on the nature and the physical state of the gas. The variation of flux by unit of surface 
$\Phi(\nu,x)$ is related to an absorption coefficient $K(\nu,x)$ :
\begin{equation}
{\rm d}\Phi (\nu,x)= \Phi (\nu,x)K(\nu,x){\rm d}x\label{eqc}\,.
\end{equation}

Equation \ref{eqb} may be integrated numerically to get $\nu$ as a function of $x$, then equation 
\ref{eqc} is integrated. 
Set $\nu_a$  the absolute frequency of a studied absorption line, $x_n$ a value of $x$ for which $H=0$ 
and $X=x-x_n$; mark a spectral element of the light by its frequency $f$ for $X=0$. Suppose now that 
the absorption is low and that the linewidth is purely Doppler, so that without a field we would have the 
absorption at a frequency $\nu$:
\begin{equation}
{\rm d}\Phi (\nu)=-\exp(-a(\nu-\nu_a))^2{\rm d}x\label{13}  
\end{equation}
equation \ref{eqb} , written with a convenient coefficient $b$ becomes, for $X$ small:
\begin{equation}
{\rm d}\nu/{\rm d}X=bX^2.
\end{equation}
Integrating
\begin{equation}
\nu=bX^3/3+f.\label{15}
\end{equation}
From equations \ref{13} and \ref{15} the variation if the intensity of the spectral element is:
\begin{equation}
{\rm d}\Phi (f)=-\exp(-a(bX^3/3+f-\nu_a)^2) {\rm d}X
\end{equation}

\begin{figure}
\begin{center}
\end{center}
\caption{ Shapes of pseudo-lines (see text). }
\end{figure}

\medskip
A numerical computation of $\Phi$ shows \cite{Moret2} that the half intensity width is nearly constant, so 
that, taking into account in the spectrum only the fast changing intensities, that is neglecting the feet of the 
lines, the pseudo lines appear as sharp as an ordinary line for the same temperature.

\section{Tentative applications to astrophysics.}
\subsection{Order of magnitude of the redshifts.}
{\it The evaluation of a redshift requires a measure or a computation of parameters which are unknown, 
difficult to measure or compute. A rough evaluation is made demonstrative, computing the density of 
molecules which would produce a shift equal to the cosmological redshift.}

\medskip
In previous papers, we had computed an order of magnitude of the redshift using equation \ref{dfin}; this 
computation required an hypothesis about the ratio of frequencies $\nu_i/F$, whose evaluation is arbitrary. 
We prefer now to use equation \ref{dafin} because it uses the index of refraction of a gas which has the 
same order of magnitude ($5.10^{-4}$ in the normal conditions) for all gases, and the Raman frequencies 
$\nu_i$, more variable, but which have often many possible values, up to the limit frequency, about 100 
MHz; the highest frequencies have the largest contribution.

We will consider that all ILSCRS active scatterings have nearly the same amplitude than the coherent 
Rayleigh scattering which produces the refraction; this hypothesis leads probably to an under-evaluation 
because, as the pulses are short, the numerous Raman oscillators are resonant with the excited oscillator, 
quickly excited.

\medskip
For a moderate redshift, Hubble's law is
\begin{equation}
\frac{\Delta\nu}{\nu L} = \frac{1}{L}\sqrt{\frac{c-LH_0}{c+LH_0}}-\frac{1}{L} \approx –
\frac{H_0}{c}\label{hubble}
\end{equation}
Combining it with equation \ref{dafin}:
\begin{equation}
\frac{c\Delta\nu}{\nu L} = H_0=(n-1)\frac{\pi h \Sigma_i \nu_i^2}{2kT}
\end{equation}
$n-1$ is proportional to the density of molecules $N$, so that, using an index 0 for the normal conditions, 
$R$ being the constant of the perfect gases:
\begin{equation}
N=N_0\frac{n-1}{n_0-1}=\frac{10^3}{22.4}\frac{2H_0RT}{\pi h (n_0-1)\Sigma_i\nu_i 
^2}=3,58.10^{35}\frac{H_0T}{(n_0-1) \Sigma_i\nu_i^2}
\end{equation}
With $H_0=2,5.10^{-18},\hskip .3em T=2,7K,\, \hskip .3em n_0-1=5.10^{-4}$ and 
$\Sigma_i\nu_i^2=10^{16}$, the number of molecules required to get the whole cosmological redshift 
from ILSCRS would be $5.10^5$ molecules per cubic meter, in the average.

The detection of the 0,21 m forbidden line shows that there is some $H_2$ molecules in the cold regions 
of the space; the UV radiation of the stars photoionizes $H_2$ into $H_2^+$ which is ILSCRS active; 
where the pressure is low so that the time between collisions is large, $H_2^+$ is stable, the proportion of 
$H_2$ molecules transformed into $H_2^+$ should be large.

\subsection{The spectra of quasars}
{\it The flexibility of ILSCRS interactions allows a lot of hypothesis \dots} 

In the spectra of the quasars, the most shifted lines are emission lines whose shifts decreases with the 
frequency; this decrease could be explained supposing that the hardest X rays, then UV, have a longer 
path in a redshifting gas. The broad absorption lines are absorbed in a colder, more external gas, thus their 
shift is lower. The emission lines may be sharp, coming from a relatively high pressure gas in which the 
redshifting is low, while the absorption is perturbed by a simultaneous frequency shift.

The Lyman absorption forest seems to be produced along the light path by intergalactic clouds generally 
close to galaxies \cite{Steidel,Schaye}; the dynamical or magnetic proposals to get these clouds thin 
enough to obtain a low Hubble line stretching do not seem to work, so that the clouds seem to be stressed 
by sheets of dark matter \cite{Croft,Blitz,Theuns}.

ILSCRS allows to consider that the clouds are thick, but that absorption leaves a visible pattern only 
where a magnetic field is nearly zero. The absorption by the remainder of the clouds is spread in the 
spectrum, it may produce a part of the absorption attributed to dust failing a better explanation 
\cite{Masci} and in despite of the difficult survival of dust \cite{Draine}.

\section{Conclusion}
The  \textquotedblleft incoherent light impulsive stimulated Raman scattering\textquotedblright plays surely 
a role in the redshift of the astrophysical spectra; it is difficult to evaluate this role precisely, but it seems 
able to help the interpretation of observations.

Other interactions of very low pressure gases with light should be studied too, for instance the cooling of 
the gases or the decrease the coefficients of absorption of the spectral lines by loss of coherence, leading 
to an under-evaluation of the gas density.

I hope that some astrophysicists will find this paper useful, allowing the introduction of interesting topics 
for their research.

\medskip
{\bf Acknowledgements}

I thank Evry Schatzman and Véronique Bommier (Paris-Meudon observatory) very much, for long and 
useful discussions.


\begin{thebibliography}{25}
\bibitem{Marmet} Marmet P., 1991 {\it Bull. of the polish Academy of Sciences} {\bf 39} 187
\bibitem{Moret} Moret-Bailly J. 1998 {\it Quantum and Semiclassical Optics} {\bf 10} L35
\bibitem{Moret1} Moret-Bailly J. 1998 {\it Ann. Phys. Fr. } {\bf 23} C1-235
\bibitem{Moret2} Moret-Bailly J. 2001 {\it J. Quantit. Spectr. \& Radiative Ttransfer} {\bf 68} 575
\bibitem{Lamb}Lamb, G. L.  Jr., 1971 {\it Rev. Mod.Phys} {\bf 43 } 99
\bibitem{Yan}Yan Y.-X., Gamble  E. B. Jr. \&  Nelson K. A. , 1985 {\it J. Chem Phys. } {\bf 83} 5391
\bibitem{Eichler} Eichler H. J., 1977 {\it Opt. Acta, 24} 631
\bibitem{Salcedo} Salcedo J. R., Siegman A. E., Dlott D. D.  \& Fayer M. D., 1978 {\it Phys. Rev. Lett. 
} {\bf 41 } 131
\bibitem{Nelson1} Nelson K. A.  \& Fayer M. D., 1980 {\it J. Chem. Phys} {\bf 72} 5202
\bibitem{Nelson2} Nelson K. A., Miller R. J., Lutz C. \& Fayer M. D., J., 1982 {\it Appl. phys} {\bf 
53} 1144
\bibitem{Robinson} Robinson M. M. R., Yan Y.-X., Gamble E. B.  Jr., Williams L. R., Meth J. S.  \& 
Nelson K. A., 1984 {\it Chem. Phys. Lett. } {\bf 112 } 491
\bibitem{De Silvestri} De Silvestri S., Fujimoto J. G., Ippen E. P., Gamble E. B. Jr., Williams L. R.  \& 
Nelson K. A., Chem. Phys. Lett {\it 1985} {\bf 116} 146
\bibitem{Bloembergen} Bloembergen N. 1982, in Nonlinear optics , Benjamin, New York,  § 2-8.
\bibitem{Webb} Webb J. K., Flambaum V. V., Churchill C. W., Drinkwater M. J. and Barrow J., 1999 
{\it Phys. Rev. Lett. } {\bf 82} 884.
\bibitem{Steidel} Steidel C. C. 1993, in QSO absorption lines, ed. G. Meylan (Springer), 139.
\bibitem{Schaye} Schaye J., Theuns T., Rauch M., Efstathiou G., \& Sargent W. L. W., 2000 {\it Mon. 
Not. R. Astron. Soc.} (to be published).
\bibitem{Croft} Croft R., Weinberg D. H., Katz N., \& Hernquist L., 1997 {\it ApJ } {\bf 488} 532
\bibitem{Blitz} Blitz L., Spergel D. N., Teuben P. J., Hartmann D., \& Burton W. B. 1999 {\it ApJ} {\bf 
514}, 818.
\bibitem{Theuns} Theuns T., Schaye J., \& Haehnelt G., 2000 {\it Mon. Not. R. Astron. Soc.} (to be 
published).
\bibitem{Masci} Masci F. J. \& Webster R. L., 1995 {\it Publ. Astron. Soc. Aust} {\bf 12} 146
\bibitem{Draine} Draine B. T. \& Salpeter E. E., 1979 {\it ApJ} {\bf 231} 77

\end{thebibliography}
\end{document}